\documentclass[a4paper,fleqn,usenatbib]{mnras}

\usepackage[T1]{fontenc}
\usepackage{ae,aecompl}

\usepackage{graphicx}	
\usepackage{amsmath}	
\usepackage{amssymb}	
\usepackage{multirow}
\usepackage{xcolor}

\newcommand{\pcc}{\,{\rm cm}^{-3}}
\newcommand{\gcc}{\,{\rm g \, cm}^{-3}}

\newcommand{\kel}{\, {\rm K}}

\newcommand{\msun}{\, {\rm M}_\odot}
\newcommand{\nh}{n_{_{\rm H}}}

\newcommand{\pc}{\, {\rm pc}}

\newcommand{\myr}{\, {\rm Myr}}

\newcommand{\kms}{\, {\rm km \, s^{-1}}}

\newcommand{\subO}{_{_{\rm O}}}

\newcommand{\subB}{_{_{\rm B}}}
\newcommand{\subC}{_{_{\rm C}}}
\newcommand{\vseq}{\!\!&\!\!=\!\!&\!\!}

\title[Isolated vs embedded cores]{Differences in chemical evolution between isolated and embedded prestellar cores}

\author[Priestley et al.]{
F. D. Priestley$^1$\thanks{Email: priestleyf@cardiff.ac.uk}, A. P. Whitworth$^1$ \& E. Fogerty$^2$
\\
$^1$School of Physics and Astronomy, Cardiff University, Queen's Buildings, The Parade, Cardiff CF24 3AA, UK \\
$^2$Center for Theoretical Astrophysics, Los Alamos National Laboratory, Los Alamos, New Mexico, USA}

\date{Accepted XXX. Received YYY; in original form ZZZ}

\pubyear{2022}

\begin{document}
\label{firstpage}
\pagerange{\pageref{firstpage}--\pageref{lastpage}}
\maketitle

\begin{abstract}

  {Models of prestellar cores often assume that the cores are isolated from their environment - material outside the core boundary plays no role in the subsequent evolution. This is unlikely to be the case in reality, where cores are located within hierarchically substructured molecular clouds. We investigate the dynamical and chemical evolution of prestellar cores, modelled as Bonnor-Ebert spheres, and show that the density of the ambient medium has a large impact on the resulting chemical properties of the cores. Models embedded in high-density, low-temperature surroundings have greatly enhanced abundances of several molecules, such as CO and CS, compared to models with more diffuse surroundings, corresponding to relatively isolated cores. The predicted intensities and profile shapes of molecular lines are also affected. The density of the ambient medium has a stronger effect on the chemical evolution than whether the cores are initially in or out of equilibrium. This suggests that the impact of environment cannot be neglected when modelling chemistry in prestellar cores; the results of these models are highly sensitive to the assumptions made about the core surroundings.}
  
\end{abstract}


\section{Introduction}

{Much of our understanding of prestellar cores is based on line emission from molecules. As well as providing kinematic information, different lines preferentially trace different parts of cores due to a combination of abundance, excitation and optical depth effects, presenting a far more detailed picture of the objects than continuum absorption or emission. This complexity also makes the interpretation of molecular line emission challenging, and theoretical models are necessary in order to fully exploit the large data sets now routinely produced by modern submillimetre facilities.}

{Prestellar cores are frequently modelled as static one-zone or spherically-symmetric structures, but the dynamical and chemical timescales in these objects are similar \citep{banerji2009}, making this approximation of decoupled physical and chemical evolution inaccurate \citep{sipila2018}. Coupling hydrodynamical simulations to a time-dependent chemical network resolves this issue \citep{aikawa2005,tassis2012,tritsis2022}, at the expense of a significant increase in complexity. Selecting an appropriate model for a particular object (or group of objects) is often done by comparing the predicted and observed line profiles \citep{rawlings1992,keto2015,sipila2018}. Models which satisfactorily reproduce the line observations (sometimes in combination with other constraints) are considered to be reasonably accurate representations of the real core/cores.}

{Most simulations of the type described above assume that cores are effectively isolated from their surroundings (although there are exceptions; \citealt{smith2012,smith2013,bovino2021}) - the core evolves independently, with outside influence limited to a confining pressure. It is not clear that this assumption is valid in reality. Prestellar cores are thought to form via supersonic turbulence within molecular clouds \citep{maclow2004}, which is difficult to reconcile with the notion of isolated evolution. Molecular clouds themselves show evidence of hierarchichal substructure, containing clumps and filaments which themselves contain cores \citep{bergin2007}, with the distinction between each level of structure being somewhat arbitrary. There are theoretical \citep{bonnell2001,clark2021} and observational \citep{peretto2020,rigby2021,anderson2021} arguments for cores continuing to accrete material from their surroundings as they evolve. Incorporating these effects is likely to significantly impact the chemical properties of model cores.}

{In this paper, we investigate differences in prestellar chemical evolution caused by a relatively benign implementation of environmental effects: embedding the model core in a dense ($\sim 100 \pcc$), cool ($\sim 10 \kel$) ambient medium, rather than isolating it completely from its surroundings. The added weight of the surrounding material changes the dynamical evolution of the core \citep{kaminski2014}, which consequently affects its chemical makeup and the predicted line profiles. We show that these differences are far from negligible, suggesting that chemical evolution in prestellar cores is strongly affected by their larger-scale surroundings.}

\begin{figure*}
  \centering
  \includegraphics[width=\columnwidth]{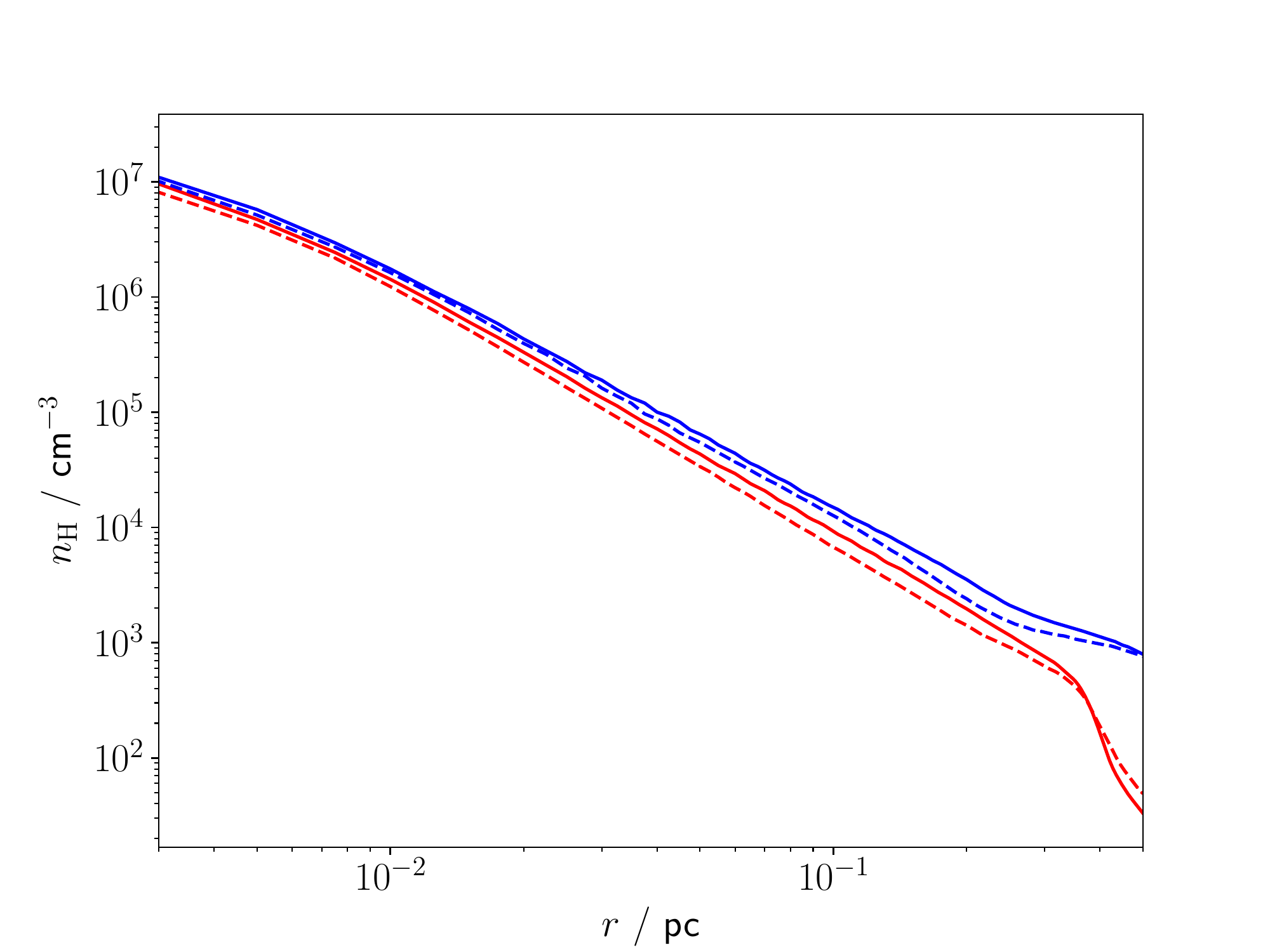}\quad
  \includegraphics[width=\columnwidth]{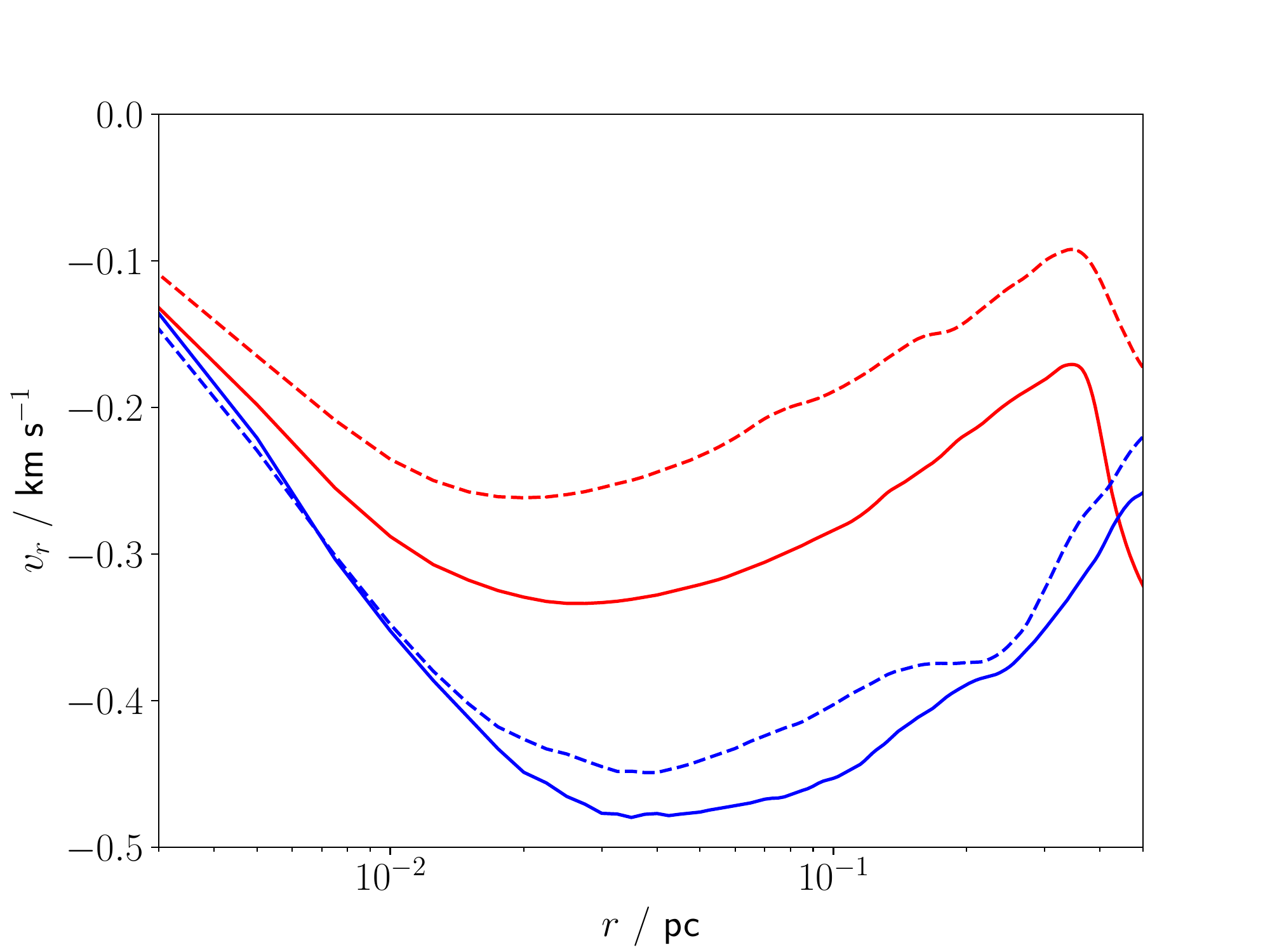}
  \caption{Density (left) and radial velocity (right) profiles at $t_{\rm end}$ for models C30F0 (red solid line), C30F1 (red dashed line), C1F0 (blue solid line) and C1F1 (blue dashed line).}
  \label{fig:profiles}
\end{figure*}

\begin{figure*}
  \centering
  \includegraphics[width=\columnwidth]{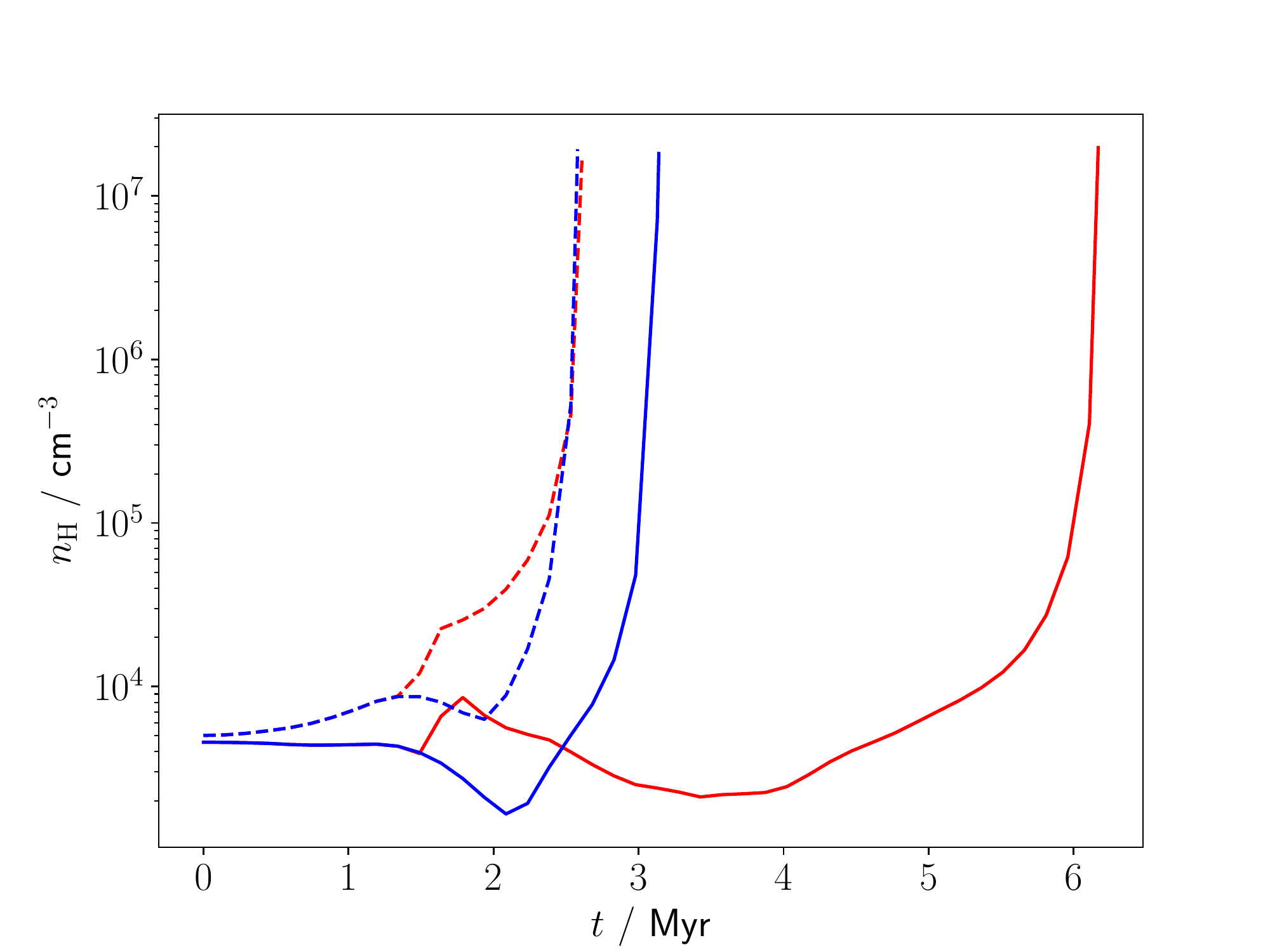}\quad
  \includegraphics[width=\columnwidth]{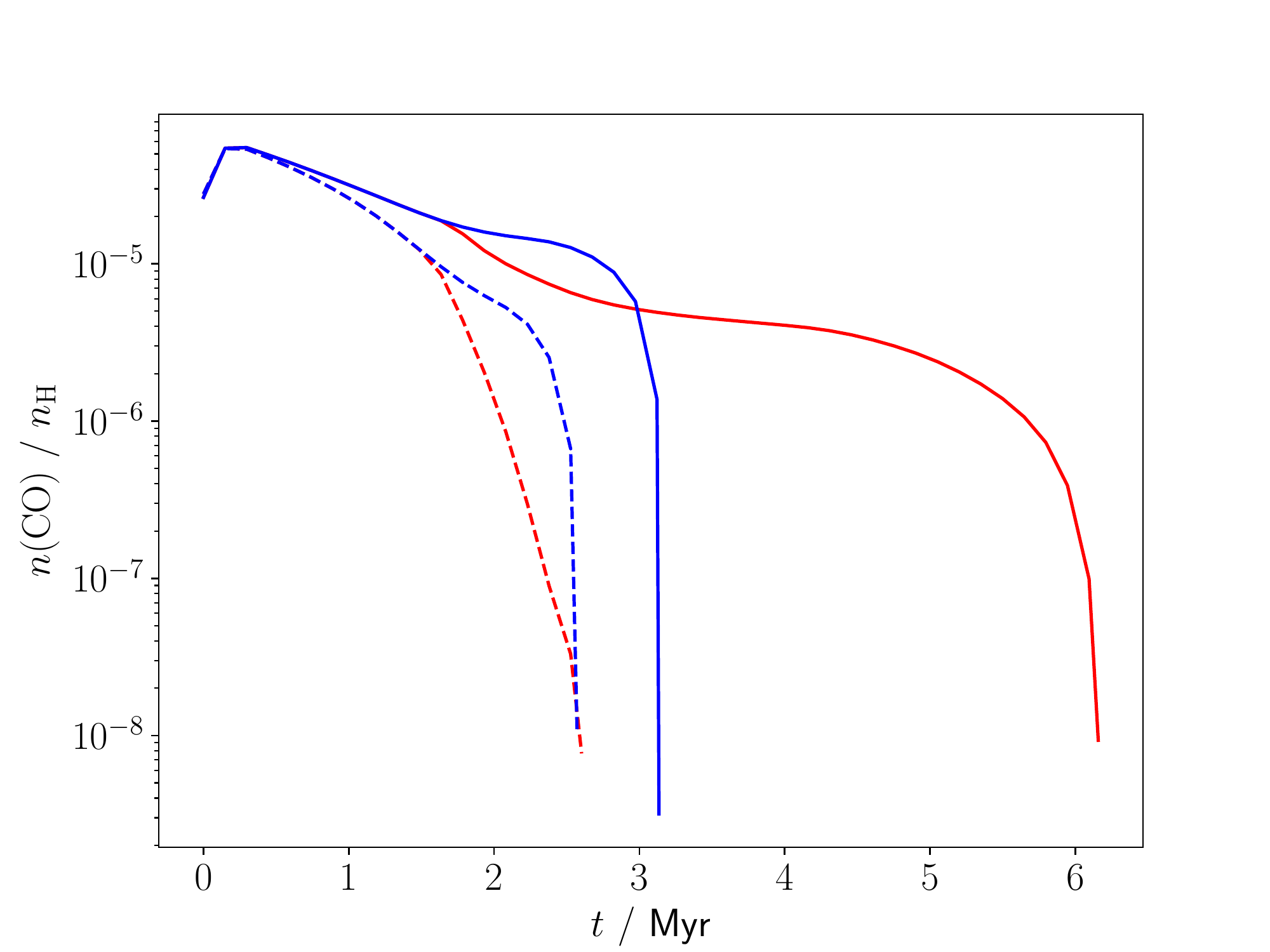}\quad
  \caption{Evolution of the central density (left) and CO abundance (right) with time for models C30F0 (red solid line), C30F1 (red dashed line), C1F0 (blue solid line) and C1F1 (blue dashed line).}
  \label{fig:rhomax}
\end{figure*}

\begin{figure*}
  \centering
  \includegraphics[width=0.66\columnwidth]{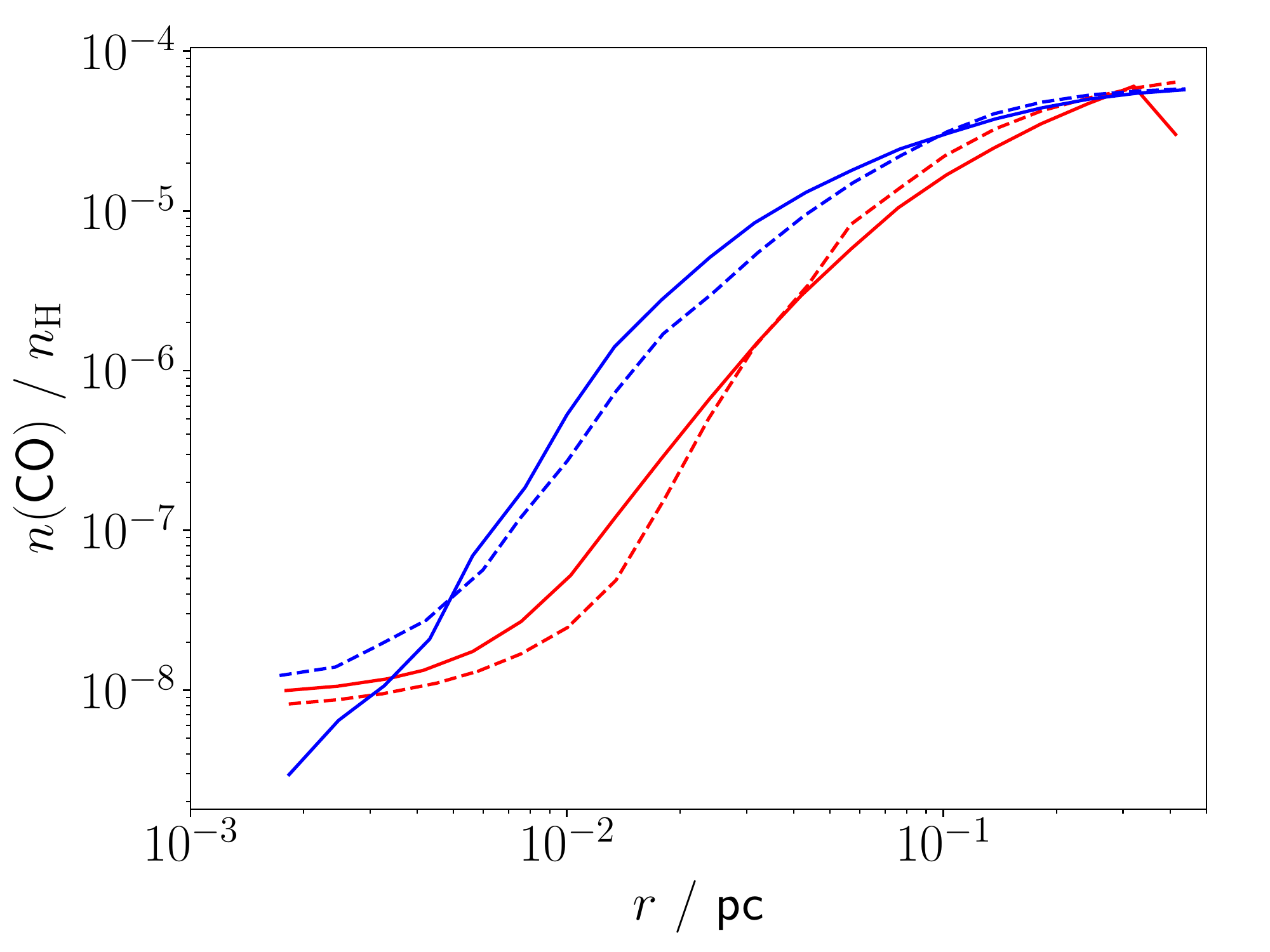}
  \includegraphics[width=0.66\columnwidth]{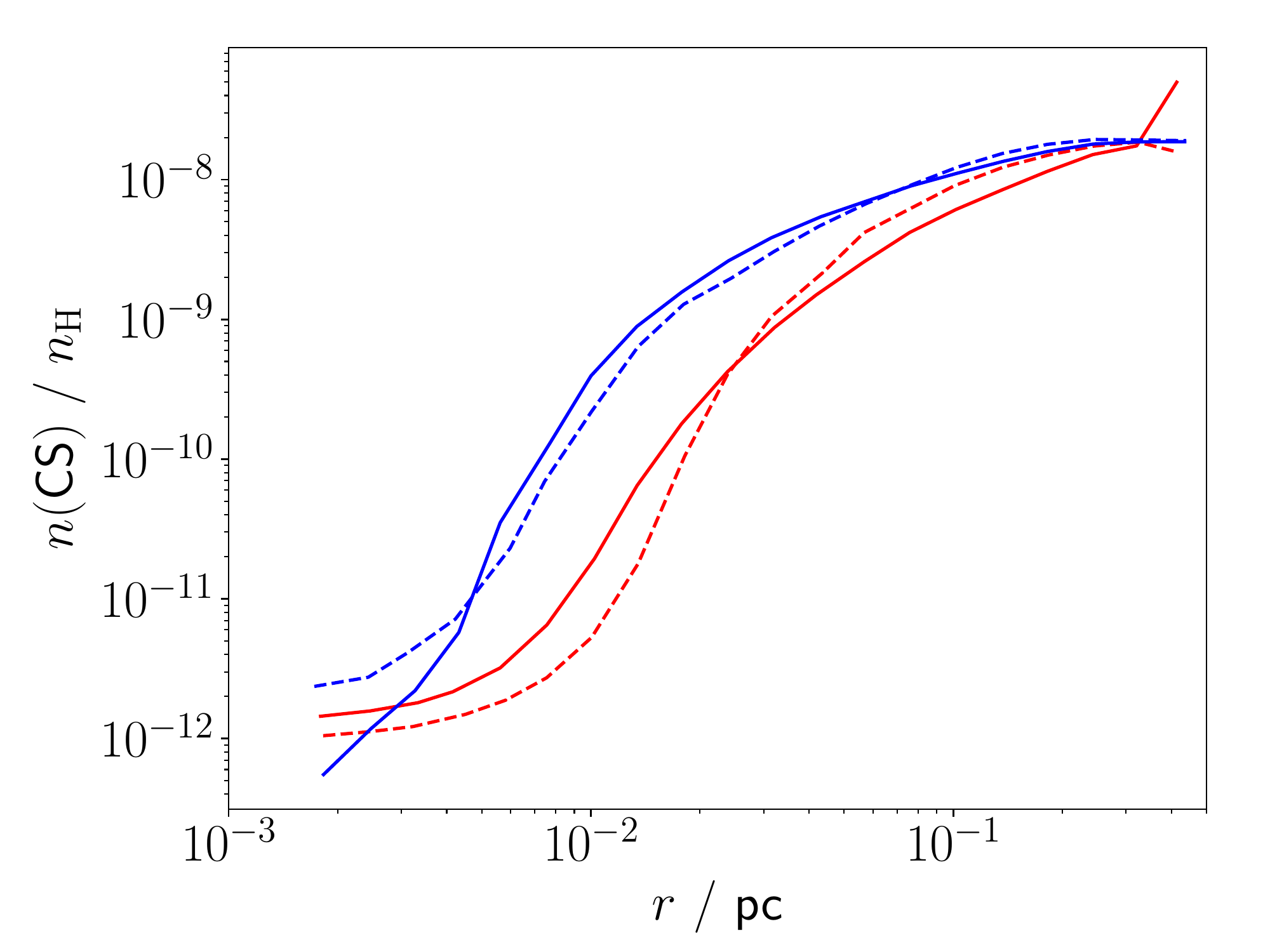}
  \includegraphics[width=0.66\columnwidth]{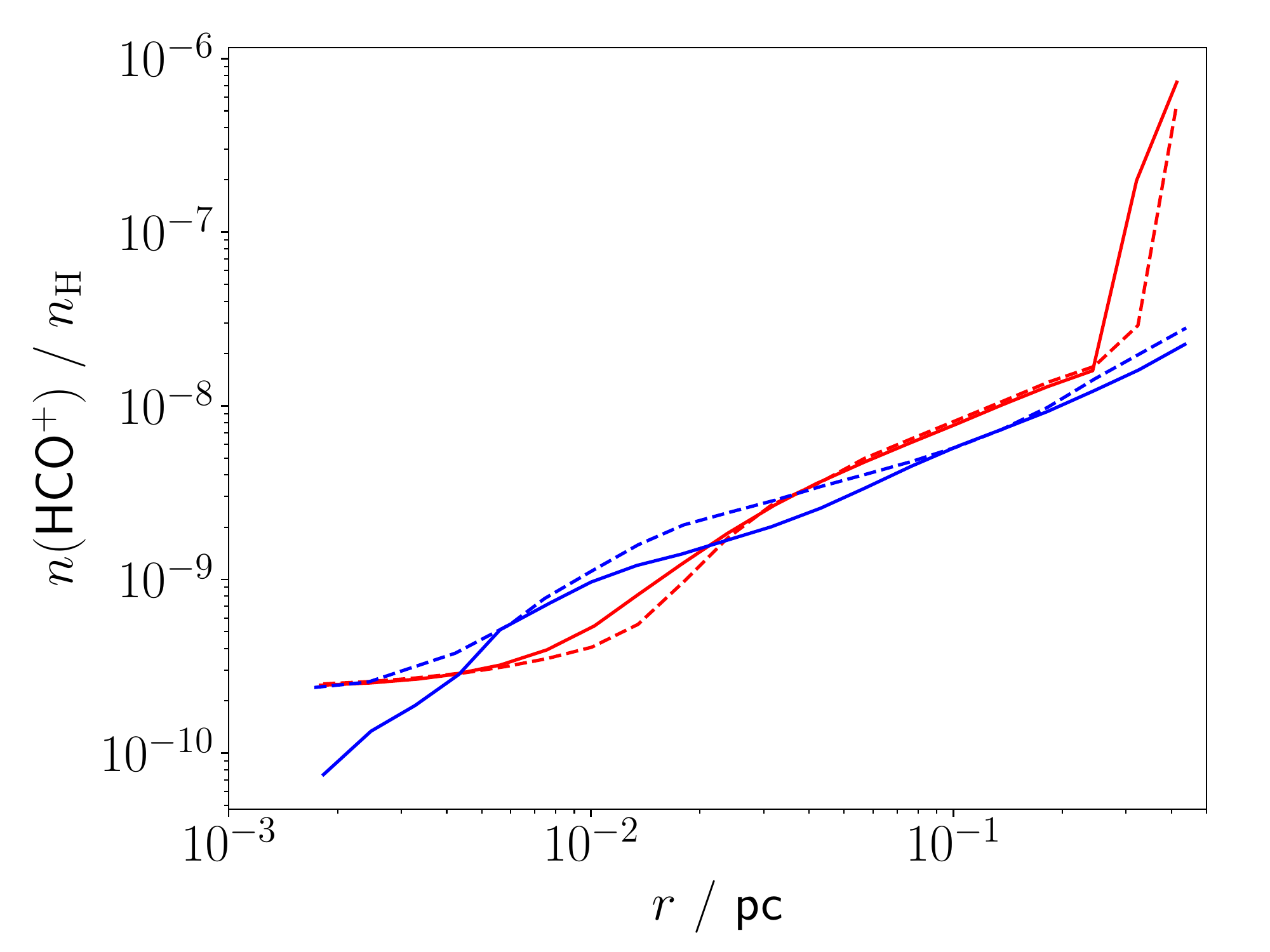}\\
  \includegraphics[width=0.66\columnwidth]{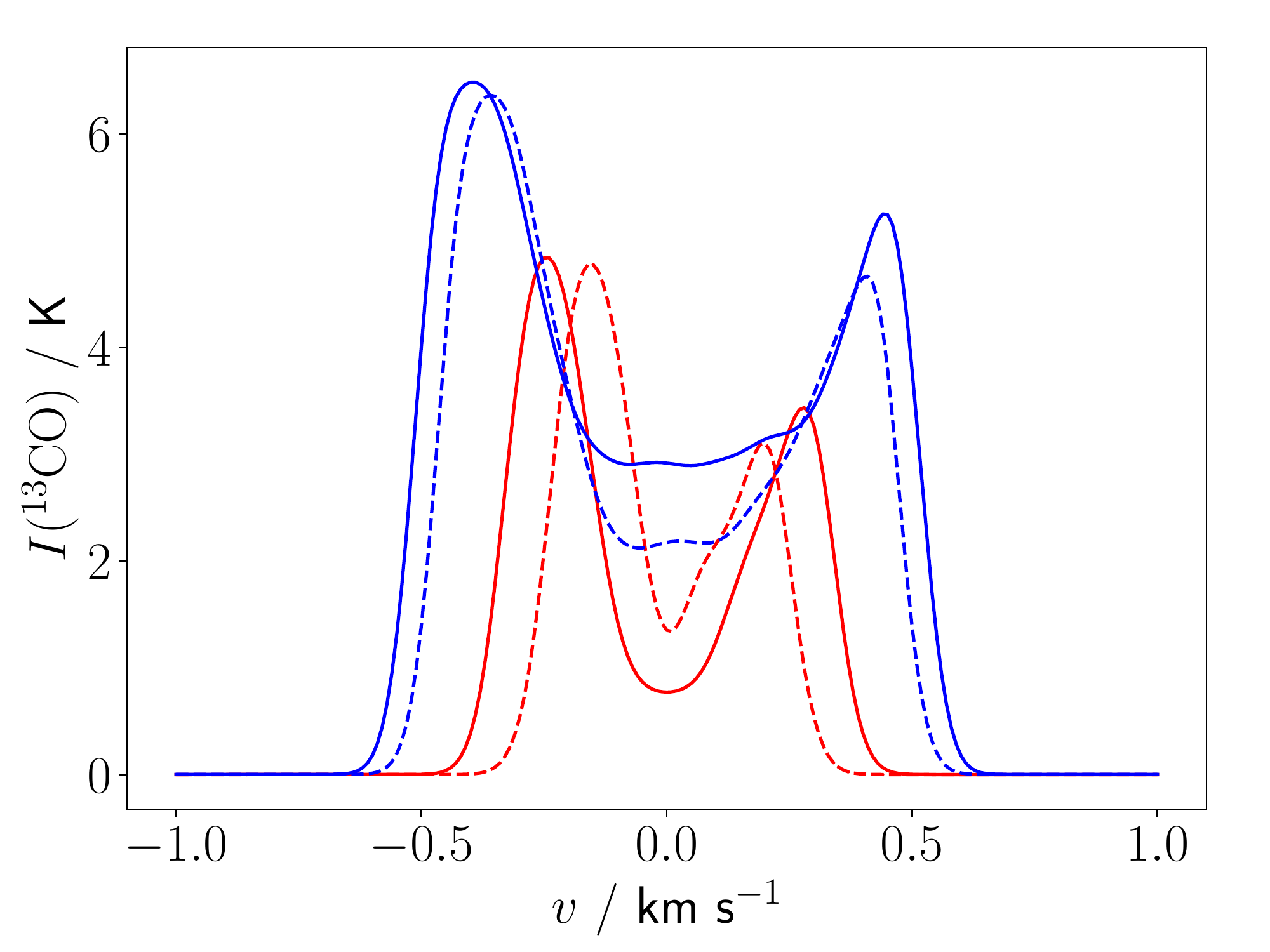}
  \includegraphics[width=0.66\columnwidth]{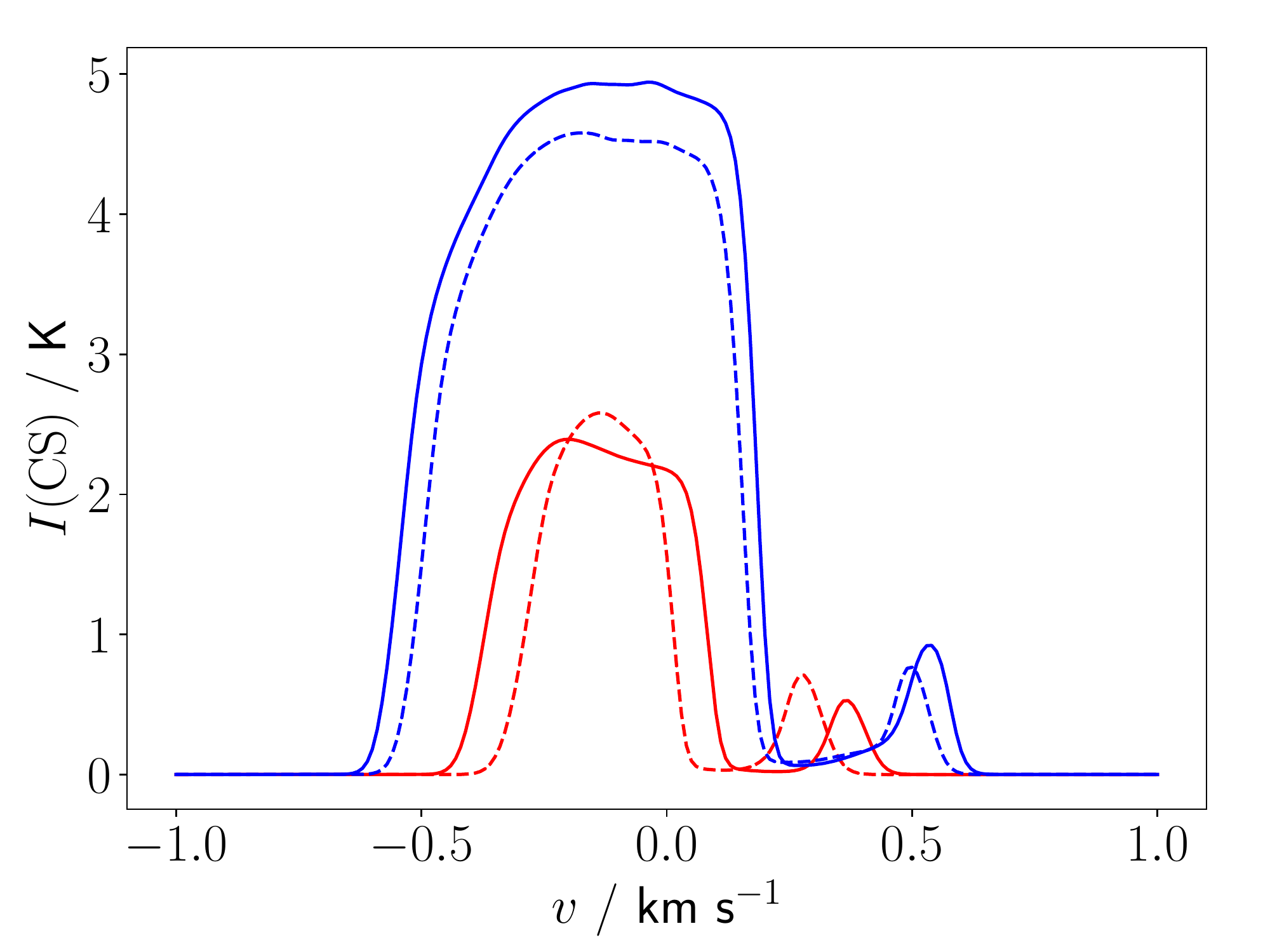}
  \includegraphics[width=0.66\columnwidth]{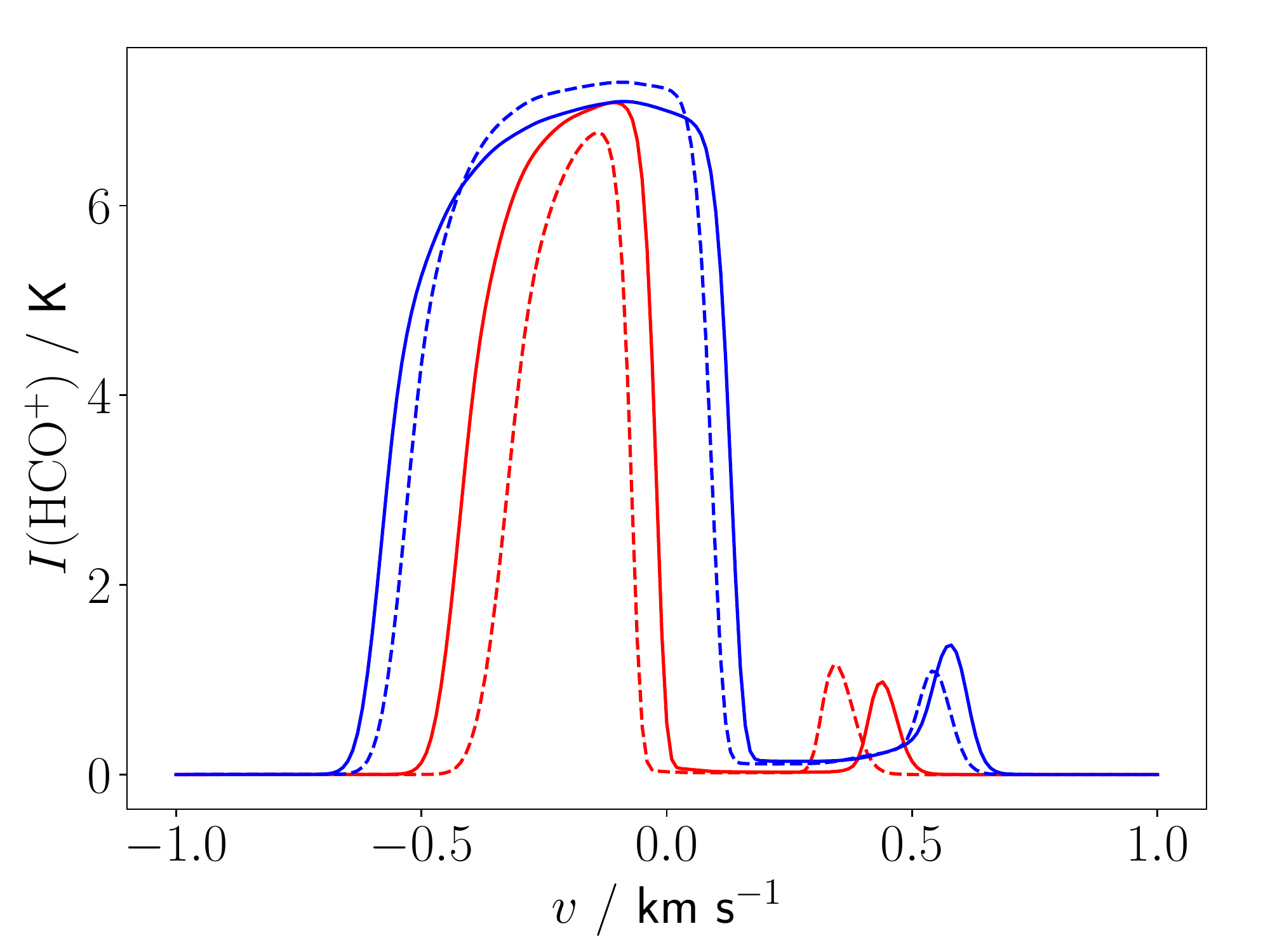}\\
  \caption{{\it Top row:} the abundance profiles for CO {\it (left panel)}, CS {\it (centre panel)} and HCO$^+$ {\it (right panel)} at $t_{\rm end}$. {\it Bottom row:} the corresponding line profiles for $^{13}$CO $J=1-0$ {\it (left panel)}, CS $J=2-1$ {\it (centre panel)} and HCO$^+$ $J=1-0$ {\it (right panel)} transitions. Results for the C30F0 model at $6.2 \myr$ are shown as red solid lines, C30F1 at $2.6 \myr$ as red dashed lines, C1F0 at $3.1 \myr$ as blue solid lines, and C1F1 at $2.6 \myr$ as blue dashed lines.}
  \label{fig:mol}
\end{figure*}

\section{Method}

\begin{table}
\centering
\caption{Model names, the values of the density contrast ${\cal C}$ and enhancement ${\cal F}$, the time at which the model reaches a central density of $10^7 \pcc$ ($t_{\rm end}$), and the time for which the central density is above $10^4 \pcc$ ($t_{\rm dense}$).}
\begin{tabular}{ccccc}
  \hline
  Model & ${\cal C}$ & ${\cal F}$ & $t_{\rm end}$ / Myr & $t_{\rm dense}$ / Myr \\
  \hline
  C30F0 & $30$ & $1$ & $6.2$ & $0.7$ \\
  C30F1 & $30$ & $1.1$ & $2.6$ & $1.1$ \\
  C1F0 & $1$ & $1$ & $3.1$ & $0.3$ \\
  C1F1 & $1$ & $1.1$ & $2.6$ & $0.3$ \\
  \hline
\end{tabular}
\label{TAB:Setups}
\end{table}

{We model prestellar cores as critical Bonnor-Ebert (BE) spheres \citep{ebert1955,bonnor1956}, spherically symmetric self-gravitating clouds of isothermal gas, in hydrostatic equilibrium and confined by an external pressure. The dimensionless radius, $\xi \equiv (4\pi G \rho\subO)^{1/2}r/a\subO$ for central density $\rho\subO$ and isothermal sound speed $a\subO$, has a critical value $\xi_{_{\rm CRIT}} \sim 6.45$ beyond which no equilibrium solutions exist. This corresponds to a physical radius $r\subB$, boundary density $\rho\subB$ and pressure $P\subB$ given by
\begin{eqnarray}
r_{_{\rm B:CRIT}}\vseq\left[4.42\!\times\!10^{-2}\,{\rm pc}\right]\!\left[\frac{M\subO}{{\rm M}_{_\odot}}\right]\!\left[\frac{a\subO}{0.2\,{\rm km/s}}\right]^{-2}\!,\\
\rho_{_{\rm B:CRIT}}\vseq\left[7.61\!\times\!10^{-20}\,{\rm g/cm^3}\right]\!\left[\frac{M\subO}{{\rm M}_{_\odot}}\right]^{-2}\!\left[\frac{a\subO}{0.2\,{\rm km/s}}\right]^6\!,\hspace{0.5cm}\\
P_{_{\rm B:CRIT}}\vseq\left[2.20\!\times\!10^5\,{\rm K/cm^3}\right]\!\left[\frac{M\subO}{{\rm M}_{_\odot}}\right]^{-2}\!\left[\frac{a\subO}{0.2\,{\rm km/s}}\right]^8\!,
\end{eqnarray}
where $M\subO$ is the mass of the sphere. We choose $M\subO = 10 \msun$ and $a\subO = 0.2 \kms$ (i.e. molecular gas at $\sim 10 \kel$), giving $r\subB = 0.44 \pc$, $\rho\subB= 7.6 \times 10^{-22} \gcc$, and $P\subB = 2.2 \times 10^3 \, {\rm K \, cm^{-3}}$.}

{We embed our BE spheres in a uniform ambient medium with density
\begin{eqnarray}\label{EQN:rhoC.01}
\rho\subC&=&{\cal C}^{-1}\,\rho\subB\hspace{1.0cm}({\cal C}\geq1),
\end{eqnarray}
and the isothermal sound-speed increased to compensate,
\begin{eqnarray}\label{EQN:aC.01}
a\subC&=&{\cal C}^{1/2}\,a\subO,\hspace{1.9cm}
\end{eqnarray}
so that the pressure in the ambient medium is equal to $P\subB$. We investigate cores with ${\cal C} = 30$, i.e. surrounded by a low-density, high-temperature ambient medium, and ${\cal C} = 1$, where the ambient medium has the same properties as the BE sphere at $r\subB$.}

{Critical BE spheres are in equilibrium, albeit an unstable one. In order to induce collapse, it is common to increase the density everywhere in the sphere by a factor ${\cal F}$ \citep{foster1993}. This increases the inward force of self-gravity by a factor ${\cal F}^{\,4/3}$ and the outward force due to the pressure gradient by a factor ${\cal F}$. Consequently hydrostatic equilibrium is perturbed in favour of self-gravity throughout the core, and it immediately starts to contract. We consider values of ${\cal F} = 1$ (quasi-equilibrium spheres) and $1.1$ (non-equilibrium spheres).}

We perform three-dimensional smoothed-particle hydrodynamic (SPH) simulations of model prestellar cores using {\sc phantom} \citep{price2018}. The SPH particle smoothing length is given by
\begin{eqnarray}\label{EQN:h.01}
h=1.2\,[m_{_{\rm SPH}}/\rho]^{1/3},
\end{eqnarray}
so each SPH particle typically has of order 57 neighbours. The computational domain is a cubic box with periodic boundary conditions, although gravity is not periodic, so that there is no gravitational attraction from outside the box.

The ambient medium has density $\rho\subC =\rho\subB/{\cal C}$, and if ${\cal C}>1$ then $\rho\subC<\rho\subB$, so there is a discontinuous density drop across the boundary of the BE sphere. However, due to the density smoothing inherent in SPH, ambient-medium particles close to this discontinuity have densities higher than $\rho\subB/C$. If these ambient-medium particles are then assigned a sound speed higher by a factor $C^{1/2}$ than those in the BE sphere, they are over-pressured, and drive a compression wave into the sphere \citep{bisbas2009}. To mitigate this problem, we adopt an equation of state
\begin{eqnarray}\label{EQN:aO.01}
  a^2(\rho) =
  \begin{cases}
    a\subO^2 \, 2^{-1/3} \left[1 + (\rho\subB/\rho)^3\right]^{1/3}, & \hfill \rho \le \rho\subB, \\
    a\subO^2\,, & \hfill \rho > \rho\subB; \\
  \end{cases}
  \label{eq:eos}
\end{eqnarray}
this has the effect of smoothing out changes in the sound speed so that the pressure in the ambient medium is everywhere exactly equal to the pressure at the boundary of the BE sphere, as shown in Appendix \ref{APP:TemperatureSmoothing}.

We use $50 \, 000$ SPH particles within the BE sphere\footnote{{We show in Appendix \ref{sec:restest} that while the simulations are not fully converged, this resolution is sufficient for our purposes.}}, and set the thickness of the ambient medium to be ten particle smoothing lengths, i.e. $10h\subB =12[{\cal C}m_{_{\rm SPH}}/\rho\subB]^{1/3} =0.20\,{\cal C}^{1/3}\pc$ (see Equation \ref{EQN:h.01}). Consequently the extent of the computational domain and the total number of SPH particles both depend on $C$. We follow the evolution of the simulated prestellar cores until the maximum density reaches $2.34 \times 10^{-17} \gcc$, corresponding to a number density of hydrogen nuclei $\nh = 10^7 \pcc$.

We randomly select $10\,000$ particles within the prestellar core, and use the time evolution of their densities to calculate the chemical evolution using {\sc uclchem} \citep{holdship2017}, a time-dependent gas-grain chemical code, and the UMIST12 reaction network \citep{mcelroy2013}. We assume gas and dust temperatures of $10 \kel$, a cosmic ray ionization rate per H$_2$ molecule of $1.3 \times 10^{-17} \, {\rm s^{-1}}$, and use the high-metal elemental abundances from \citet{lee1998}. {We assume the cores are sufficiently well-shielded that we can ignore the effect of any external ultraviolet (UV) radiation on the chemistry. This is not necessarily the case in real prestellar cores, but as we are primarily interested in differences between models rather than the results of the individual models themselves, any external UV field should have little impact on our conclusions.} Cosmic ray-generated UV photons are included.

As molecular abundances are not directly observable quantities, we also calculate emission line profiles for our models using {\sc lime} \citep{brinch2010}. We take molecular properties from the LAMDA database \citep{schoier2005} and dust optical properties from \citet{ossenkopf1994}. We use $10\,000$ sample points, and assign each one the properties of the closest chemically post-processed SPH particle. We adopt a $^{12}$C/$^{13}$C ratio of 77 from \citet{wilson1994}. Line profiles are extracted from the central $0.1 \pc$ of the model, corresponding to typical beam sizes of single-dish observations of nearby cores \citep[e.g.][]{tafalla2002}.

{We consider four models, with parameters listed in Table \ref{TAB:Setups}, for the four combinations of ${\cal C} = 30/1$ and ${\cal F} = 1/1.1$. This allows us to distinguish effects which are due to the ambient density from those due to whether the core is initially in or out of equilibrium. ${\cal C} = 30$ models are our approximation to truly isolated (${\cal C} = \infty$) cores, as the ambient medium has relatively little effect (although it will still eventually perturb the core into collapse). ${\cal C} = 1$ models represent cores embedded within molecular clouds. Observable differences between truly isolated cores and embedded ones should be at least as large as those between models with ${\cal C} = 1$ and $30$.}

\section{Results}

{Figure \ref{fig:profiles} shows the density and velocity profiles of the models at $t_{\rm end}$, when the central density has reached $10^7 \pcc$. All four models have very similar final density profiles - as noted by \citet{kaminski2014}, they tend to evolve towards the same late-time behaviour. The velocity profiles are also similar in shape. Howeber, the magnitude of the radial velocity differs substantially between models, and the evolutionary pathways to the final state are very different.}

{Figure \ref{fig:rhomax} shows the time evolution of the central density and central CO abundance. Models with a density enhancement (${\cal F} = 1.1$) begin contracting from the outset, and approach freefall collapse at slightly less than one sound-crossing time ($t_{\rm SC} = r\subB/a\subO = 2.2\,{\rm Myr}$). Both models terminate at $2.6 \myr$, and evolve similarly except for during a brief period between $1.5$ and $2 \myr$ where the central density and CO abundance tracks diverge.}

{For ${\cal F} = 1$, the cores are initially in quasi-equilibrium, oscillating about the initial state, but the weight of the ambient medium pushes a compression wave\footnote{This is distinct from the artificial compression waves discussed in Appendix \ref{APP:TemperatureSmoothing}, particularly as the C1F0 model has no density or temperature discontinuity, and so the effect discussed in Appendix \ref{APP:TemperatureSmoothing} does not occur.} into the core, which eventually provokes collapse. The timescale on which this occurs depends on the density contrast, taking roughly one sound-crossing time for ${\cal C} = 1$ and two for ${\cal C} = 30$. The C30F0 simulation thus has a significantly longer duration than the other three, with the C1F0 model having a comparable $t_{\rm end}$ to the non-equilibrium ones. The C30F0 model's evolution resembles a more stretched-out version of the C1F0 model. For higher values of ${\cal C}$, the increase in timescale grows until the compression wave becomes dynamically irrelevant \citep{kaminski2014}.}

{The rapid increase in central density during the freefall phase causes a corresponding rapid decline in CO abundance due to freeze-out onto grain surfaces, for all four models. However, for ${\cal C} = 30$, this decline is more gradual than for ${\cal C} = 1$, due to the weaker compression wave caused by the ambient medium. The ${\cal C} = 30$ cores thus spend a significantly longer period of time at the high densities where freeze-out is effective; the C30F0 and C30F1 models have a central density above $10^4 \pcc$ for $0.7$ and $1.1 \myr$ respectively, compared to only $0.3 \myr$ for both C1F0 and C1F1.}

{While all four models eventually reach similar {\it central} CO abundances, the reduced window of opportunity for freeze-out in the ${\cal C} = 1$ models results in higher CO abundances throughout the rest of the core, as shown in Figure \ref{fig:mol}. At a radius of $0.01 \pc$, the CO abundance is around a factor of $100$ higher than in the ${\cal C} = 30$ models. Molecules susceptible to freeze-out, such as CS, show similar behaviour to CO, whereas others like HCO$^+$ are mostly unaffected by differences in the collapse dynamics. We note that while the molecular abundances are sensitive to the value of ${\cal C}$, the value of ${\cal F}$ has a fairly minor effect - cores which are initially (quasi)stable are chemically indistinguishable from unstable cores, despite having very different evolutionary histories. The critical factor is the strength of the compression wave driven by the ambient medium, and its impact on the timescale of the late-time density evolution.}

{The differences in the molecular abundances result in corresponding differences in the line emission from the model cores, shown in Figure \ref{fig:mol}. Again, models with the same ${\cal C}$ but different ${\cal F}$ appear very similar, with the most significant differences being between ${\cal C} = 30$ and ${\cal C} = 1$. The $^{13}$CO $J=1-0$ line has a high enough optical depth that the large differences in abundance do not result in large changes in the line intensity, but the line profile is much broader for ${\cal C} = 1$, and the shape of the self-absorption feature is different from the ${\cal C} = 30$ models. The CS $J=2-1$ line, on the other hand, has a peak intensity in the ${\cal C} = 1$ models twice that of the ${\cal C} = 30$ ones, in addition to being broader and having much stronger self-absorption. The intensity of the HCO$^+$ $J=1-0$ line is not greatly affected, as the abundances are comparable for all models, but it is again significantly broader for ${\cal C} = 1$.}

\section{Discussion}

{Our results in Figure \ref{fig:mol} suggest that the environment of prestellar cores has a significant impact on their chemical and observational properties. This effect is rarely accounted for (or even considered) in simulations, which is a concern if these simulations are to be used to interpret observational data. As an example, if one wanted to determine the age of a core by finding the point in the simulations where the predicted line profiles best match those observed, the result would clearly depend on whether the core is modelled as an isolated or embedded object, and on the properties of the ambient medium in the embedded case. We note that the density profiles for all four models (Figure \ref{fig:profiles}) are so similar as to be effectively indistinguishable, so they cannot be used to discriminate between the different scenarios.}

{While we have only investigated the simple case of a uniform and static ambient medium, there is almost no limit in principle to the density and velocity fields that one could choose to embed model cores inside, making the predictions of chemical evolution models basically unconstrained. Trying to come up with specific models to represent individual cores seems to us to be futile under these circumstances; one can never be sure whether conclusions drawn from the models apply in general, or only for some specific choice of background. We suggest that chemical studies should explore the properties of large samples of model cores, formed self-consistently from simulations of larger-scale structures \citep{smith2012,smith2013,bovino2021}, or covering a wide range of the parameter space of initial conditions \citep{priestley2022}. Consistent trends in molecular properties identified in this way might be considered more robust to details of the core environment, and therefore useful for interpreting observations of real cores.}

Our line profiles in Figure \ref{fig:mol} are somewhat broader, and have stronger self-absorption features, than those typically seen in observations of prestellar cores \citep[e.g.][]{lee1999,tafalla2002}, {regardless of the ambient density or equilibrium state of the core. Previous studies \citep{keto2015,sipila2018} have obtained narrower line profiles, closer to those observed, for cores with more extended initial density profiles, which reach their final central densities with mostly-subsonic infall velocities. These models have initial dimensionless radii greater than the critical value for stability ($\xi = 15$ in \citealt{sipila2018}, compared to $\xi_{_{\rm CRIT}} \sim 6.45$), and forming such an unstable object in the dynamic environment of a turbulent molecular cloud is not obviously feasible (\citealt{whitworth1996}, although see discussion in \citealt{keto2015}). \citet{yin2021} find that introducing a dynamically-important magnetic field results in better agreement between modelled and observed line profiles (see also \citealt{priestley2022}), although this is in tension with measured field strengths in prestellar cores \citep{pattle2022}. Our results here suggest that the impact of the environment on line profiles can be comparable to these other effects, which may provide an alternative resolution to the problem.}

\section{Conclusions}

We have simulated the chemical evolution of prestellar cores, modelled as BE spheres, {varying the initial equilibrium state and the density contrast with the ambient medium. We find that cores with the same density contrast appear chemically similar, regardless of whether they are initially quasistable or out of equilbrium, despite very different evolutionary timescales. Changing the density contrast, on the other hand, results in significant changes to both the molecular abundances and the predicted line profiles. This is due to the varying strength of the compression wave driven into the core by the weight of the surrounding material, suggesting that the environment of prestellar cores has a non-trivial impact on their chemical evolution. Many studies neglect the existence of any material outside the core boundary, and those which do not rarely consider it to be an important aspect of the model. We argue that this is unrealistic; the evolution of cores cannot be considered in isolation, and the results of models which do not address this should be treated with caution.}

\section*{Acknowledgements}

FDP and APW gratefully acknowledge the support of a Consolidated Grant from the UK Science and Technology Facilities Council (ST/K00926/1).

\section*{Data Availability}

The data underlying this article will be made available on request.

\bibliographystyle{mnras}
\bibliography{bechem}

\appendix

\section{Temperature smoothing across the boundary of the BE sphere}\label{APP:TemperatureSmoothing}

\begin{figure}
  \includegraphics[width=\columnwidth]{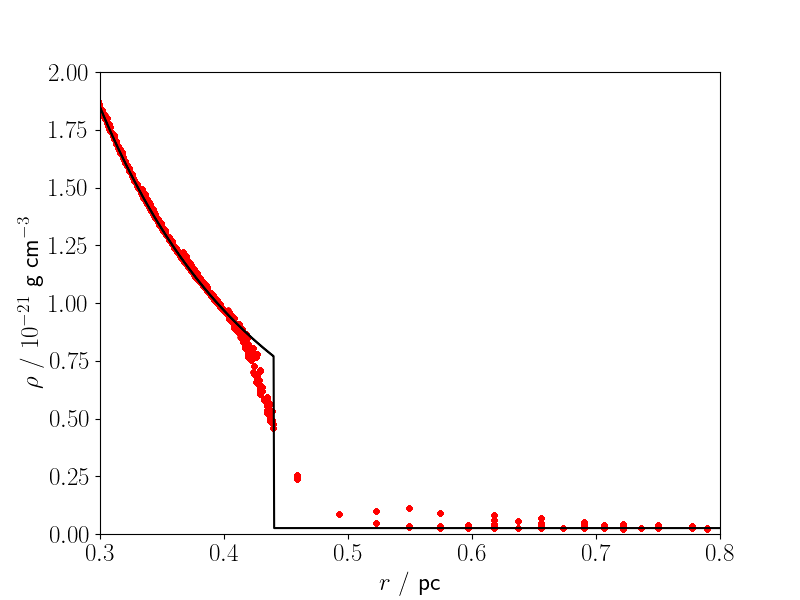}\\
  \includegraphics[width=\columnwidth]{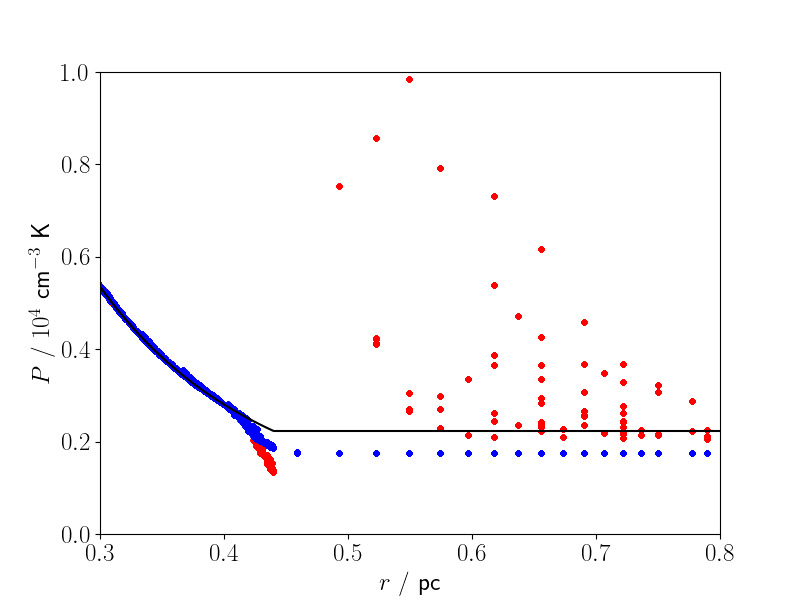}
  \caption{{\it Top panel:} Exact initial density profile of the C30F0 model (black line), and the actual particle densities as a function of radius (red points). {\it Bottom panel:} Exact thermal pressure profile for the C30F0 model (black line), and particle thermal pressures for i) an isothermal equation of state, with the sound speed increased by a factor of ${\cal C}^{1/2}$ for particles with $r > r_{_{\rm B:CRIT}}$ (red points); and ii) the equation of state given by Equation \ref{eq:eos} (blue points). Note that we show only a small fraction of the BE sphere near the boundary.}
  \label{fig:smooth}
\end{figure}

{The top panel of Figure \ref{fig:smooth} shows the density profile of the C30F0 model, calculated by integrating the isothermal equation up to $r_{_{\rm B:CRIT}}$ and setting the density to ${\cal C}^{-1}\,\rho\subB$ beyond that point, compared to the actual SPH particle densities as a function of radius. The particle densities follow the exact solution throughout most of the core, but near the boundary, the SPH smoothing kernel starts to include the lower-density ambient medium, and the model density is lower than it should be. Similarly, there are regions of the ambient medium with a higher density than the nominal ${\cal C}^{-1}\,\rho\subB$. When combined with an isothermal equation of state, with the sound speed modified according to Equation \ref{EQN:aC.01}, this gives rise to a discontinuity in the pressure profile, as shown in the bottom panel of Figure \ref{fig:smooth}: rather than being in pressure equilibrium, points inside the BE sphere near the boundary are underpressurised compared to the lower-density points just outside the boundary. This drives an artificial compression wave into the BE sphere, affecting its subsequent evolution.}

Using the smoothed equation of state given by Equation \ref{eq:eos}, this bump disappears, and pressure equilibrium is restored across the boundary of the BE sphere. The pressure does not match the exact solution, due to the factor of $2^{-1/3}$ in Equation \ref{eq:eos}, but the gradients in the profile are removed, preventing (or at least mitigating) unphysical behaviour. We note that the width of the region in which the temperature smoothing is important decreases with the number of SPH particles used in the BE sphere. However, although an increase in the number of SPH particles improves the accuracy of the simulation, the need for temperature smoothing does not go away.

\section{Resolution tests}
\label{sec:restest}

\begin{figure*}
  \centering
  \includegraphics[width=0.66\columnwidth]{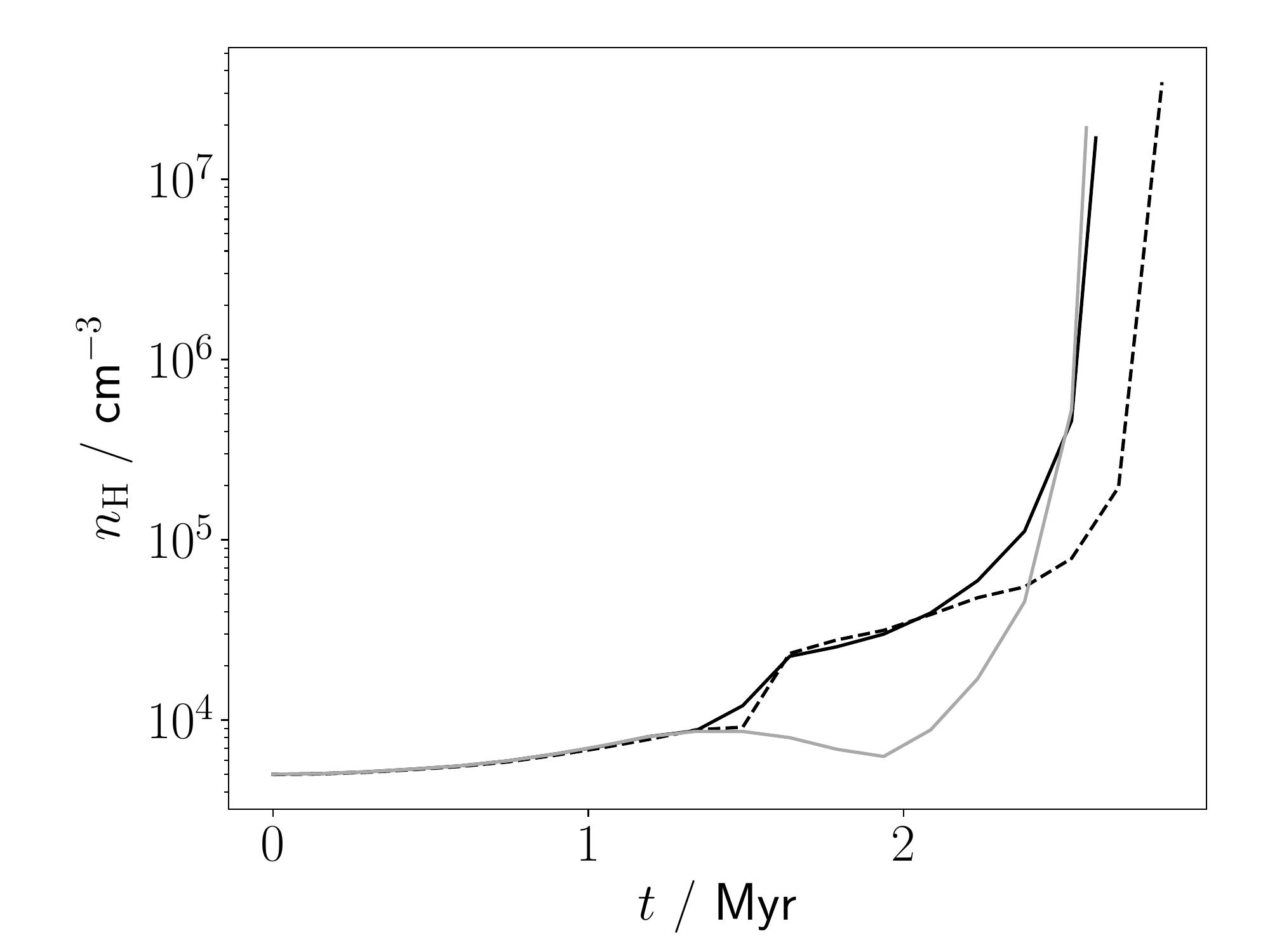}\quad
  \includegraphics[width=0.66\columnwidth]{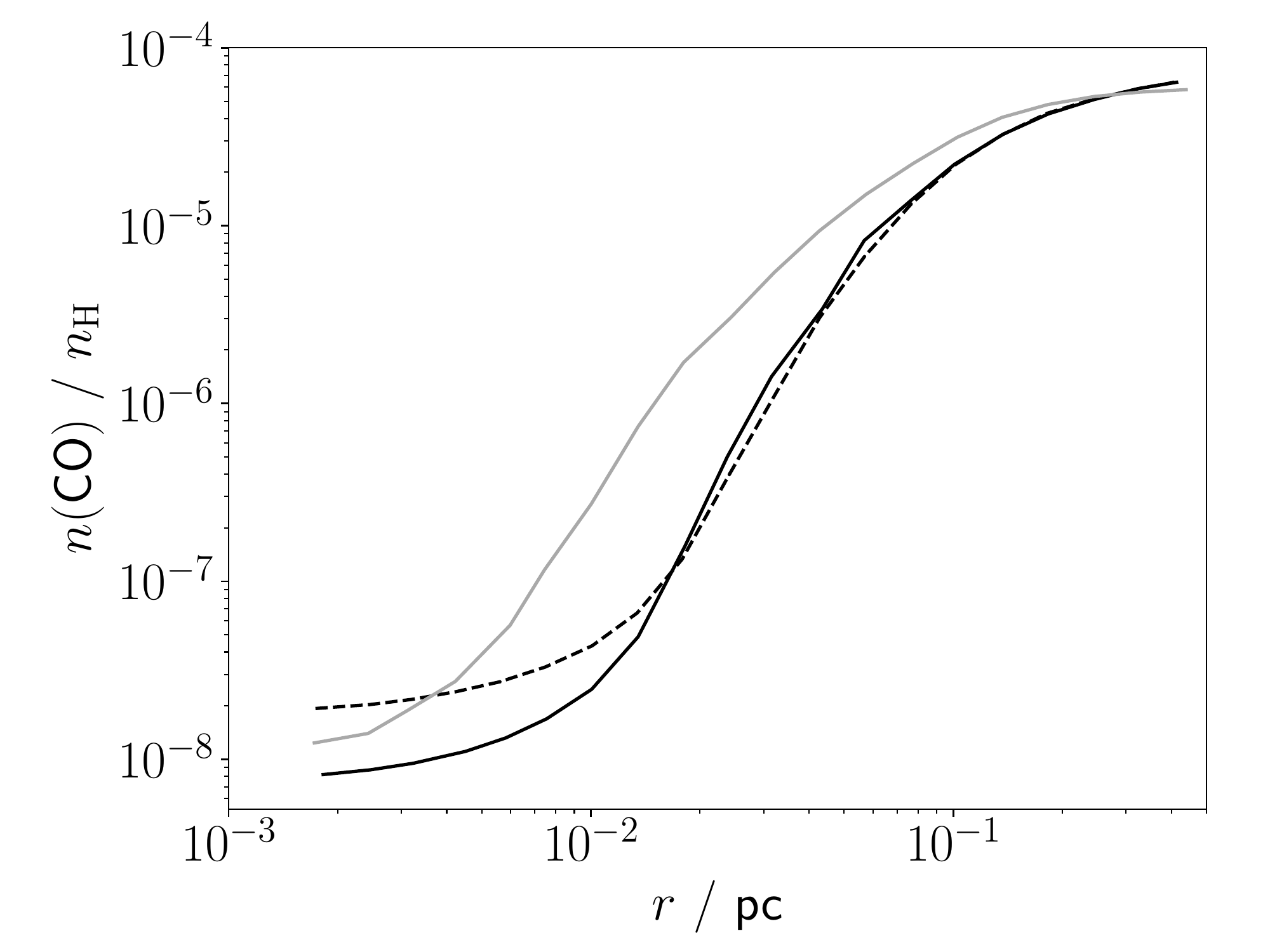}\quad
  \includegraphics[width=0.66\columnwidth]{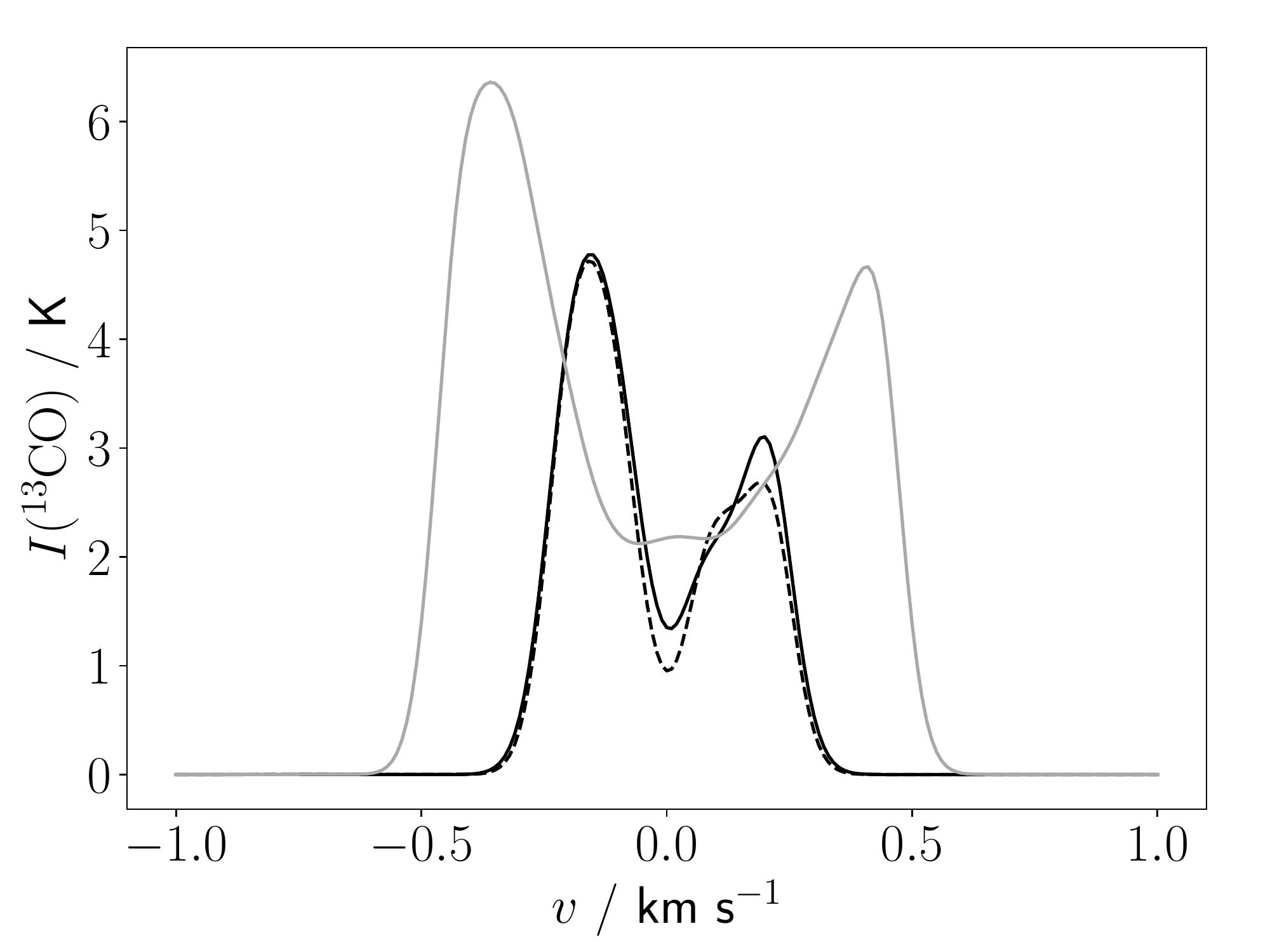}\quad
  \caption{Evolution of the central density (left), CO abundance profiles at $t_{\rm end}$ (middle), and $^{13}$CO line profiles (right), for the C30F1 model with standard (solid black lines) and enhanced (dashed black lines) resolution. The C1F1 model with standard resolution is shown for comparison (solid grey lines).}
  \label{fig:restest}
\end{figure*}

{Figure \ref{fig:restest} shows the effects of increasing the resolution by a factor of two for all stages of the modelling procedure for the C30F1 model, i.e. $100 \, 000$ SPH particles in the BE sphere, of which $20 \, 000$ are post-processed chemically, and $20 \, 000$ {\sc lime} sample points. The hydrodynamical simulations are not fully converged with $50 \, 000$ particles - the high-resolution C30F1 model reaches the final density of $10^7 \pcc$ at $2.8 \myr$, rather than $2.6 \myr$ for the standard resolution, and spends $1.2$ rather than $1.1 \myr$ above a central density of $10^4 \pcc$. However, the overall evolution is still very similar to the standard resolution case, and as such, differences in the molecular abundances and line profiles due to the increased resolution are minor compared to the differences caused by changing the background density. Higher resolution in the C30F1 model causes a slight increase in central CO abundance, and almost no change in the $^{13}$CO line profile. The C1F1 model (with standard resolution) is still clearly distinct from both cases.}

\bsp	
\label{lastpage}
\end{document}